\documentclass{ws-procs9x6}
\begin{document}

\title{
Separatrix chaos: new approach to the theoretical treatment}

\author{S. M. Soskin
}
\address{Institute of Semiconductor Physics, Pr. Nauki 45,\\
Kiev, 03028, Ukraine\\
E-mail: smsoskin@g.com.ua}
\address{Abdus Salam ICTP, Strada Costiera 11,\\
Trieste, 34100, Italy\\
E-mail: ssoskin@ictp.it}

\author{R. Mannella}
\address{Dipartimento di Fisica, Universit\`{a} di Pisa, Largo Pontecorvo 3,\\
Pisa, 56127, Italy\\
E-mail: mannella@df.unipi.it}

\author{O. M. Yevtushenko}
\address{Physics Department,
   Ludwig-Maximilians-Universit{\"a}t M{\"u}nchen\\
M{\"u}nchen, D-80333, Germany}
\address{Abdus Salam ICTP, Strada Costiera 11,\\
Trieste, 34100, Italy\\
E-mail: bom@ictp.it}

\begin{abstract}
We develop a new approach to the theoretical treatment of the
separatrix chaos, using a special analysis of the separatrix map.
The approach allows us to describe boundaries of the separatrix
chaotic layer in the Poincar\'{e} section and transport within the
layer. We show that the maximum which the width of the layer in
energy takes as the perturbation frequency varies is much larger
than the perturbation amplitude, in contrast to predictions by
earlier theories suggesting that the maximum width is of the order
of the amplitude. The approach has also allowed us to develop the
self-consistent theory of the earlier discovered (PRL 90, 174101
(2003)) drastic facilitation of the onset of global chaos between
adjacent separatrices. Simulations agree with
the theory.
\end{abstract}
\bodymatter

\section{Introduction}

Even a weak perturbation of an integrable system possessing a
separatrix results in the onset of chaotic motion inside a
layer\cite{Zaslavsky:1991,lichtenberg_lieberman,zaslavsky:1998,zaslavsky:2005,treshev}
which we shall further call as the separatrix chaotic layer (SL).
The separatrix chaos plays a fundamental role for the Hamiltonian
chaos, being also relevant to various
applications\cite{Zaslavsky:1991,lichtenberg_lieberman,zaslavsky:1998,zaslavsky:2005,treshev}.
The boundaries of the SL in the Poincar\'{e} section may be easily
found numerically\cite{zaslavsky:1998}. However, it is also
important, both from the theoretical and practical points of view,
to be able to {\it theoretically} calculate them and describe
transport within the SL.

One of the most powerful theoretical tools for the SL study is the
separatrix map (SM), introduced in \cite{ZF:1968} for the nearly
integrable systems with the 3/2 degrees of freedom and called
sometimes\cite{treshev} as the Zaslavsky separatrix map. It may also
be generalized for systems with more degrees of freedom and for
strongly non-integrable systems (see  \cite{treshev} for the most
recent major review). We shall further consider the case of the 3/2
degrees of freedom but the generalization of our method for other
cases may be done too.

One of the most interesting for physical applications relevant
quantities is the width of the SL in energy \cite
{Zaslavsky:1991,lichtenberg_lieberman,zaslavsky:1998,zaslavsky:2005,ZF:1968,soskin2000,PR,prl2003,pre2007,2}.
There are various heuristic criteria
\cite{Zaslavsky:1991,lichtenberg_lieberman,zaslavsky:1998,zaslavsky:2005}
based on the separatrix map and various conjectures. The width by
these criteria does not depend on the angle and, as a function of a
perturbation frequency $\omega_f$, possesses a maximum at
$\omega_f$ of the order of the eigenfrequency in the stable state
$\omega_0$ while the maximum itself is of the order of the
perturbation amplitude $h$. However, the work \cite{prl2003} has
demonstrated in simulations for double-separatrix systems that the
maximum width may be {\it much larger} as the SL absorbs one or two
{\it nonlinear resonances}. The recent work\cite{pre2007} has proved
this, developing a new method for the analysis of the separatrix
map. The method is of a {\it general} validity, as shown in the {\it
present} work. We show that the maximum width occurs at the
frequency which is typically smaller than $\omega_0$ by the
logarithmic factor $\ln(1/h)$ while the maximum width is typically
much larger than $h$ - either by a numerical factor or by the
logarithmic factor (apart from the adiabatic divergence in certain
class of systems\cite{2}). Besides, the method allows to describe
major statistical properties of transport within the SL.

Note that there were various mathematical works considering the SL
in rather different contexts (see \cite{treshev} for the review). In
particular, they analysed the SL width in
normal coordinates. However, to the best of our knowledge, these
works do not specify the relation between the normal coordinates and
variables conventional in physics (e.g. energy-angle or
coordinate-momentum). Besides, these works just estimate the width
from above and below while our method allows to carry out an {\it
accurate} calculation of the width in energy and, moreover, of the
SL boundaries in the Poincar\'{e} section. Finally and most
importantly, the methods described in \cite{treshev} do not make a
resolution between the resonance frequency range and other frequency
ranges while our method shows that the SLs in these ranges
drastically differ from each other, as confirmed by simulations.

Below, we describe the basic ideas of our method (Sec. 2), review
the results of its application to the double-separatrix case (Sec.
3) and present rough estimates for the single-separatrix case (Sec.
4).

\section{Basic ideas}

Consider any 1D Hamiltonian system possessing at least one
separatrix. Let us add a weak time-periodic perturbation,

\begin{equation}
H=H_0(p,q) +hV(p,q,t),\quad V(p,q,t+2\pi/\omega_f)=V(p,q,t),\quad
h\ll 1.
\end{equation}

The motion near any of the separatrices
may be approximated by the separatrix map (SM) \cite
{Zaslavsky:1991,lichtenberg_lieberman,zaslavsky:1998,zaslavsky:2005,treshev,ZF:1968,pre2007}.
The map slightly differs for different types of separatrix. Our
method applies to all types but, to be concrete, we consider in this
section only the separatrix
with  a single saddle and two loops (like in a double-well potential
system). Then the SM reads as \cite{pre2007} (cf. also \cite
{Zaslavsky:1991,lichtenberg_lieberman,zaslavsky:1998,zaslavsky:2005,treshev,ZF:1968}):

\begin{eqnarray}
E_{i+1}&=&E_i+\sigma_ih\epsilon\sin(\varphi_i),
\\
\varphi_{i+1}&=&\varphi_i+\frac{\omega_f\pi(3- {\rm
sign}(E_{i+1}-E_s)) }{2\omega(E_{i+1})}, \nonumber
\\
\sigma_{i+1}&=&\sigma_i \, {\rm sign}(E_s-E_{i+1}), \nonumber
\\
&&\epsilon \equiv \epsilon(\omega_f)= -{\rm sign}(\left.\partial
H_0/\partial p\right|_{t\rightarrow -\infty})
\int_{-\infty}^{\infty}
{\rm d}t\;\left.\partial H_0/\partial p\right|_{E_s}\sin(\omega_ft)
,\nonumber
\\
&&E_i\equiv \left.H_0(p,q)\right|_{t_i}, \quad\varphi_i\equiv
\omega_ft_i,\quad\sigma_i\equiv{\rm sign}(\left.\partial
H_0/\partial p\right|_{t_i}), \nonumber
\end{eqnarray}

\noindent where $E_s$ is the separatrix energy. The quantity
$\epsilon$ is often called as the
Melnikov\cite{lichtenberg_lieberman} or
Poincar\'{e}-Melnikov\cite{treshev} integral. The quantity
$\delta\equiv h|\epsilon|$ is sometimes called {\it separatrix
split}\cite{zaslavsky:1998}. For the sake of simplicity, let
absolute values of all parameters of $H_0$ and of $V$ be $\sim 1$.
Then $|\epsilon|\sim 1$ too, if $\omega_f\stackrel{<}{\sim}1$.

Consider two most general ideas of the method.

{\bf 1.} The SM trajectory that includes any state with $E=E_s$ is
{\it chaotic} since the angle of this state is not correlated with
the angle of the state at the preceding step of the map, due to the
divergence of $\omega^{-1}(E\rightarrow E_s)$.

{\bf 2.} The frequency of eigenoscillation as a function of energy
is proportional to the reciprocal of the logarithmic factor:

\begin{eqnarray}
&&\omega(E)=\frac{a\pi\omega_0}{\ln\left(\Delta H/|E-E_s|\right)},
\quad \quad a=\frac{3- {\rm sign}(E-E_s)}{2},
\\
&& |E-E_s|\ll\Delta H\sim E_s-E_{st}^{(1)}\sim
E_s-E_{st}^{(2)}\nonumber
\end{eqnarray}

\noindent ($E_{st}^{(1,2)}$ are energies of the stable states).

Given that the argument of the logarithm is large in the relevant
range of $E$, the function $\omega (E)$ is nearly constant at a
rather significant variation of the argument. Therefore, as the SM
maps the state $(E_0=E_s,\varphi_0)$ onto the state with
$E=E_1\equiv E_s+\sigma_0 h\epsilon\sin(\varphi_0)$, the value of
$\omega(E)$ for the given ${\rm
sign}(\sigma_0\epsilon\sin(\varphi_0))$ is nearly the same for most
of angles $\varphi_0$ (except the close vicinity of $\pi$
multiples), namely

\begin{equation}
\omega(E)\approx\omega_r^{(\pm)}\equiv\omega(E_s\pm
h
)\quad\quad {\rm for} \quad {\rm
sign}(\sigma_0\epsilon\sin(\varphi_0))=\pm 1.
\end{equation}

Moreover, even if the deviation of the trajectory of the SM from the
separatrix further increases/decreases, $\omega(E)$ remains close to
$\omega_r^{(\pm)}$ provided the deviation is not too large/small,
namely if $|\ln((E-E_s)/h)|\ll\ln(\Delta H/h)$. If
$\omega_f\stackrel{<}{\sim}\omega_r^{(\pm)}$, then the evolution of
the map (2) may be regular-like for a long time until the energy
returns back to the close vicinity of the separatrix, where the
correlation of angle is lost again. Such a behavior is especially
pronounced if the perturbation frequency is close to
$\omega_r^{(+)}$ or $\omega_r^{(-)}$ or to one of their multiples of
a not too high order: the resonance between the perturbation and
eigenoscillation gives rise to an accumulation of the energy gain
for many steps of the SM, which results in the deviation of $E$ from
$E_s$ greatly exceeding the separatrix split $\delta\sim h$. As a
function of $\omega_f$, the largest (along the SL boundary)
deviation from the separatrix takes its maximum at the frequencies
slightly exceeding $\omega_r^{(+)}$ and $\omega_r^{(-)}$, for the
upper and lower boundaries of the SL respectively. This corresponds
to the absorption of the nonlinear resonance by the SL.

The description of the regular-like parts of the chaotic trajectory
in the case close to the resonance may be done within the resonance
approximation. The explicit matching between the SM and the
resonance approximation is carried out in \cite{pre2007}. The
resonance approximation is done in terms of slow variables, action
$I\equiv I(E)$ (note that ${\rm d}I/{\rm d}E=\omega^{-1}(E)$) and
slow angle $\tilde{\psi}\equiv\psi-\omega_f t\equiv\psi-\varphi$, by
means of the resonance Hamiltonian $\tilde{H}(I,\tilde{\psi})$
\cite{Zaslavsky:1991,lichtenberg_lieberman,zaslavsky:1998,zaslavsky:2005,PR,prl2003,pre2007}.

\begin{figure}
\psfig{file=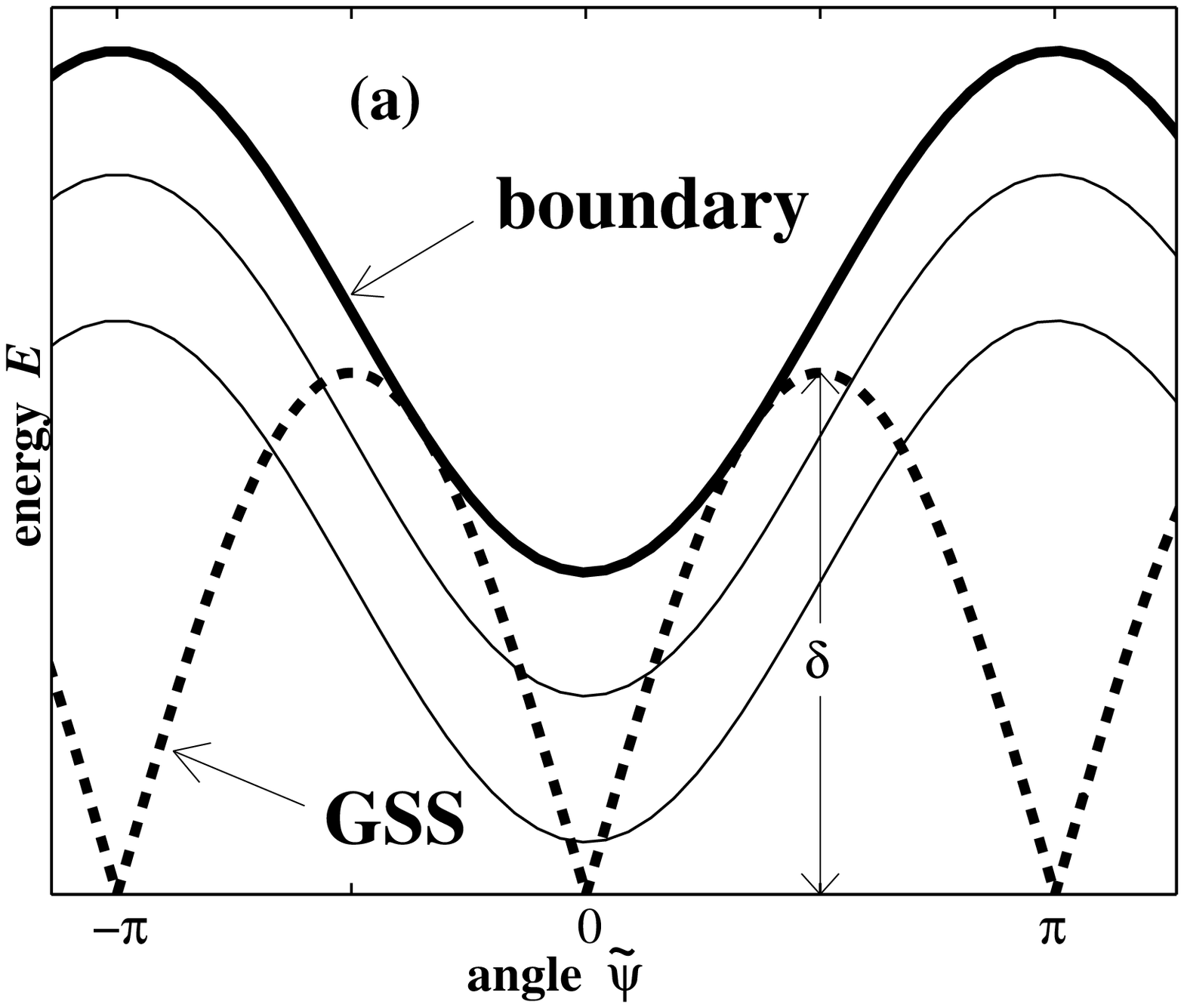,width=2.1in}
\psfig{file=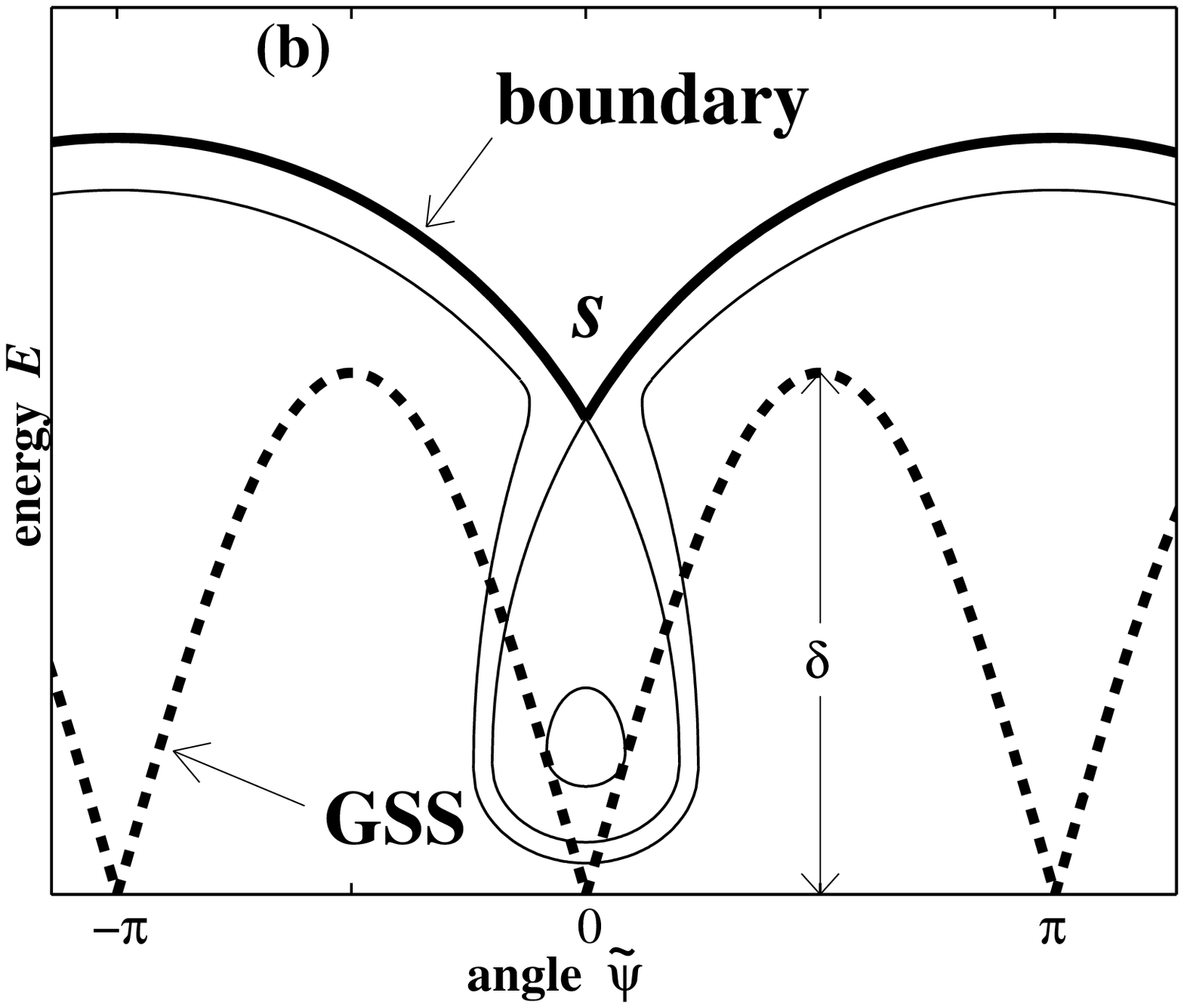,width=2.1in} \caption{Schematic illustrations
to the formation of the SL upper boundary. The separatrix energy
$E_s$ corresponds to the lower boundary of the box. The GSS curve is
shown by the dashed line. Solid lines show examples of resonant
trajectories (RTs) overlapping the GSS curve. The SL boundary (thick
solid line) is formed by: (a) the RT {\it tangent} to the GSS curve,
or (b) the upper part of the {\it self-intersecting} RT (resonant
separatrix).} \label{aba:fig1}
\end{figure}
Fig. 1 schematically illustrates the formation of the upper boundary
of the SL in the Poincar\'{e} section presented in the
$E-\tilde{\psi}$ variables. The chaotic trajectory jumps from the
separatrix onto the curve which we call the upper {\it generalized
separatrix split} (GSS) curve, within an even $\pi$ interval:

\begin{eqnarray}
&& E_{\rm GSS}^{(+)}(\tilde{\psi})=E_s+\delta|\sin(\tilde{\psi})|,\quad\quad \delta\equiv h|\epsilon|,\\
&& \tilde{\psi}\in [\pi+2\pi n, 2\pi+2\pi n],\quad\quad n=0,\pm 1,
\pm 2, ...\nonumber
\end{eqnarray}

\noindent The GSS curve relates to the SM equation for $E$ with
$E_0=E_s$. The relevance of just the even $\pi$ intervals of $\tilde
{\psi}$ is a consequence of a necessarily {\it positive} sign of
$E_1-E_0$ as far as we consider the {\it upper} boundary of the SL.
Then the system follows the trajectory of the resonance Hamiltonian
$\tilde{H}$ ($E$ necessarily increases initially because
$\tilde{\psi}(0)$ is within an even $\pi$ interval), until it again
reaches the GSS curve (necessarily in an odd $\pi$ interval, where
$E$ decreases). After that, the system jumps onto the separatrix,
where the angle correlation is lost. There is a statistical
distribution of regular-like parts of the chaotic trajectory
corresponding to the homogeneous distribution of initial angles. The
boundary of the SL is formed by the trajectory of the resonance
Hamiltonian (RT) starting from the GSS curve at such initial angle
that provides for the deviation of the RT
from the separatrix to be larger than that for any other initial
angle. From the topological point of view, there may be two
different situations: either the relevant RT is {\it tangent} to the
GSS curve (Fig. 1(a)) or it intersects it while being a
self-intersecting trajectory i.e. a {\it separatrix} of the
resonance Hamiltonian (Fig. 1(b)). In the latter case, the {\it
outer} boundary is formed by the {\it outer} part of the separatrix
(in Fig. 1(b), relating to the upper boundary, this part lies above
the saddle $s$; an example of the boundary formed by the separatrix
of a different type is shown in Fig. 4(a)).

The boundary strongly depends on the angle. The maximal/minimal
deviation of $E$ from $E_s$ is much larger/smaller than $\delta$.

\begin{figure}
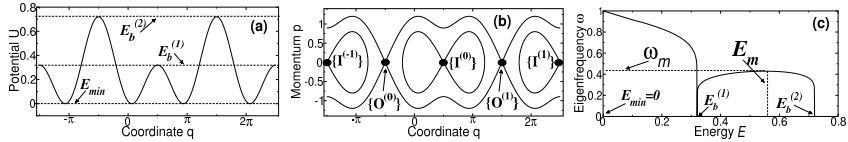

\psfig{file=Fig2a.eps,width=1.43in}
\psfig{file=Fig2b.eps,width=1.43in}
\psfig{file=Fig2c.eps,width=1.43in} \caption{The potential
$U(q)=(0.2-\sin(q))^2/2$, the separatrices and the eigenfrequency
$\omega(E)$ of the unperturbed system $H_0=p^2/2+U(q)$, in (a), (b)
and (c), respectively.} \label{aba:fig2}
\end{figure}

\begin{figure}
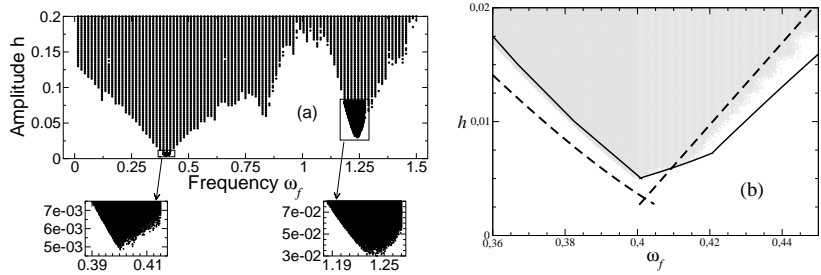

\psfig{file=Fig3a.eps,width=2.2in} \hskip 0.1in
\psfig{file=Fig3b.eps,width=1.9in} \caption{(a). The diagram
indicating (shading) the perturbation parameters range for which
global chaos exists in the perturbed system,
$H=H_0(p,q)-hq\cos(\omega_ft)$. (b). The comparison of the major
minimum of the diagram with the lowest-order theory (dashed lines)
and the theory allowing for higher-order corrections (solid lines).}
\label{aba:fig3}
\end{figure}

\section{Application to the double-separatrix chaos}

It has been found in \cite{prl2003} that, if the unperturbed
Hamiltonian $H_0$ possesses more than one separatrix (cf. Fig. 2)
while the perturbation is time-periodic, the onset of global chaos
in between the adjacent separatrices possesses a remarkable feature:
it is greatly facilitated if the perturbation frequency $\omega_f$
is close to certain frequencies: the perturbation amplitude $h$
required for the chaos onset is much smaller for such frequencies
than for others (Fig. 3). This is related
to the characteristic shape of $\omega(E)$ in between the
separatrices: $\omega(E)$ approaches a {\it rectangular} form in the
asymptotic limit of a small separation between the separatrices (cf.
Fig. 2(c)).
If  $\omega_f$ is slightly smaller than the local maximum $\omega_m$
of  $\omega(E)$, then there are two nonlinear resonances that are
very wide in energy: they may simultaneously  overlap each other and
the separatrix chaotic layers, that occurs at the value of $h$ which
is logarithmically smaller than a typical value required for the
chaos onset when $\omega_f$ lies beyond the immediate vicinity of
$\omega_m$.

Using the semi-heuristic approach, the
lowest-order asymptotic theory based on the resonance Hamiltonian
analysis was developed in \cite{prl2003} for the minima of the
global chaos boundary $h_{gc}(\omega_f)$. The theory was compared to
results of computer simulations for the given example (for a
moderately small separation: Fig. 2). The value of $\omega_f$
in the
minimum was well described by the lowest-order formula but the
descrepancy for $h_{gc}$ in the minimum was nearly 50$\%$.
Besides, it
was unclear how the overlap of resonances with the chaotic layers
occured and why even a small excess of $h$ over $h_{gc}$ resulted in
the onset of chaos in a large area of the Poincar\'{e} section
despite that chaotic layers associated with the nonlinear resonances
were exponentially narrow for $h\leq h_{gc}$. These problems have
been resolved in our recent work\cite{pre2007} developing the method
similar to that described in the previous section. The agreement
between the theory and simulations has greatly improved (Fig. 3(b)).

Our present work generalizes the above method for {\it any
separatrix} (see Sec. 2 above and Sec. 4 below). The general
validity of the method is brightly demonstrated by Fig. 4, that
shows the direct comparison of the theoretical SL boundaries and the
SLs generated by computer. Though the layers are still related to
the system with two separatrices, the given perturbation amplitude
is so small that the layers are well separated from each other and
therefore the presence of a second separatrix does not play a
significant role for any of the layers. The lower and upper layers
demonstrate two characteristic situations: the lower layer has not
yet absorbed the relevant nonlinear resonance (though the closeness
to the resonance gives rise to a rather strong increase of the layer
width) while the upper layer has absorbed the resonance so that its
maximal width greatly exceeds that of the lower layer and there is a
large island of stability in the layer.

\begin{figure}
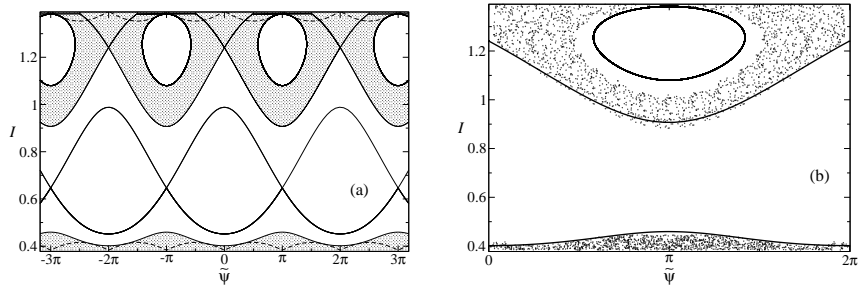

\psfig{file=Fig4a.eps,width=2.1in} \hskip 0.2in
\psfig{file=Fig4b.eps,width=2.1in} \caption{The separatrix chaotic
layers (shaded) in the plane of action $I$ and slow angle
$\tilde{\psi}$ for the system exploited in Fig. 3, for $h=0.003$ and
$\omega_f=0.401$, as described by our theory. The dashed lines
represent the relevant GSS curves. The solid lines represent
relevant trajectories of the resonance Hamiltonian $\tilde{H}$: two
solid lines with the saddles represent separatrices of the nonlinear
resonances, two other solid lines are the trajectories of
$\tilde{H}$ which are tangent to the lower and upper GSS curves
respectively. (b). Comparison of the layers obtained from computer
simulations (dots) with the theoretically calculated boundaries
(solid lines) shown in the box (a).} \label{aba:fig4}
\end{figure}

\section{Single-separatrix layer: estimates of the largest width
}

As already mentioned, the SL width in energy takes its maximum at
$\omega_f\approx\omega_r^{(+/-)}$. The rigorous treatment for the
single-separatrix case may be done similar to the double-separatrix
case \cite{pre2007}. It will be done elsewhere while, here, we
present rough estimates for the width.

Let us transform from $p-q$ to action $I\equiv I(E)$ and angle
$\psi$ and expand
$V$ in Eq. (1) into the double Fourier series (in
$t$ and
$\psi$) \cite{Zaslavsky:1991}:

\begin{equation}
V=\frac{1}{2}\sum_{k,l}V_{k,l}{\rm e}^{\imath
(k\psi-l\omega_ft)}+c.c.
\end{equation}

Let us single out the term with the maximum absolute value of
$V_{k,l}$ (typically, it corresponds to $k=l=1$) and denote it as
$V_0{\rm e}^{\imath \theta}$:

\begin{equation}
V_0(I)=\max_{k,l}(|V_{k,l}|).
\end{equation}

The maximum width of the SL corresponds to the perturbation
frequency at which the SL has just absorbed the widest {\it
nonlinear resonance} (cf. \cite{pre2007}). From the rigorous results
for the double-separatrix case\cite{pre2007}, we may assume that the
SL width is dominated by the width of the nonlinear resonance then,
i.e. the width of the resonance separatrix in energy $\Delta E$
greatly exceeds the minimal separation in energy between the
resonance separatrix and $E_s$. Obviously, this assumption should be
verified in the end.

Strictly speaking, $V_0$ strongly depends on $I$ in the relevant
range of $I$, end the rigorous treatment of the nonlinear resonance
is complicated (cf. \cite{PR,prl2003,pre2007}). But for a rough
estimate of the resonance width, it is sufficient to use a simple
Chirikov approximation of the resonance Hamiltonian
$\tilde{H}(I,\tilde{\psi})$
\cite{Zaslavsky:1991,lichtenberg_lieberman,zaslavsky:1998,zaslavsky:2005,PR,Chirikov:79},
which reduces to the auxiliary pendulum dynamics. The width of the
corresponding resonance separatrix in energy is
expressed as

\begin{equation}
\Delta E\approx\sqrt{\frac{8hV_0\omega_f}{|{\rm d}\omega/{\rm
d}E|}}.
\end{equation}

The Chirkov approximation assumes that $|{\rm d}\omega/{\rm
d}E|\approx {\rm const}$ within the range of energies relevant to
the resonance separatrix. In our case, it is not so since the
quantity $\left|{\rm d}\omega/{\rm d}E\right
|\approx\omega_f^2/(a\pi\omega_0|E-E_s|)$ strongly varies within the
relevant range of energies. However, we may still use Eq. (8) for
the rough estimate, putting in it $|E-E_s|\sim\Delta E$. Then we
obtain for $\Delta E$ the following rough asymptotic
equation

\begin{equation}
\Delta E\sim V_0(E=E_s\pm\Delta E)h\ln(1/h),\quad\quad h\rightarrow
0,
\end{equation}

\noindent where we took into account that the relevant $\omega_f$ is
close to $\omega_r^{(+/-)}\sim \omega_0/\ln(1/h)$ and omitted
numerical factors.

The asymptotic solution of Eq. (9) essentially depends on the
function $V_0(x)$. In this context, all perturbed systems may be
divided in two classes.

{\bf 1.} The separatrix of the unperturbed system has {\it more than
one saddle} while the relevant coefficient $\tilde{V}_{l}\equiv
\tilde{V}_{l}(E,\psi)$ in the Fourier expansion (in time) of the
perturbation $V$ possesses {\it different} values in adjacent
saddles. An archetypal example is a pendulum subject to a dipole
time-periodic perturbation \cite{Zaslavsky:1991,2}. If $E$ is close
to $E_s$, then the system stays mostly near one of the saddles, so
that $\tilde{V}_{l}$ depends on $\psi$ nearly in a piece-wise
manner: it oscillates between the values corresponding to the
adjacent saddles. Therefore, $V_0$ (which is the absolute value of
the relevant coefficient in the Fourier expansion of $\tilde{V}_{l}$
in $\psi$) approaches in the asymptotic limit $h\rightarrow 0$ some
{\it non-zero} constant. As follows from Eq. (8), $\Delta E$ is {\it
logarithmically large}:

\begin{equation}
\Delta E\sim h\ln(1/h)\gg h,\quad\quad h\rightarrow 0.
\end{equation}

\noindent This estimate agrees with the rigorous result and the
result of simulations in the case considered in \cite{pre2007}.

{\bf 2.} Either the separatrix has a {\it single} saddle (like for a
double-well potential system \cite{treshev,soskin2000}) or the
separatrix has more than one saddle while the perturbation possesses
{\it identically equal} values at different saddles. Archetypal
examples are a pendulum in a wave with the same spatial period
\cite{Zaslavsky:1991,lichtenberg_lieberman,zaslavsky:1998,zaslavsky:2005}
and a pendulum with the oscillating suspension point \cite{treshev}.
Then $\tilde{V}_{l}(E\rightarrow E_s, \psi)$ keeps nearly one and
the same value for most of the period of $\psi$ (as it stays most of
the period near the saddle/s): it significantly differs from this
value only during a small part of the period, which is $\sim
\omega(E)/\omega_0$. Hence, $V_0(E=E_s\pm\Delta E)\sim
1/\ln(1/\Delta E)$,
so that the solution of Eq. (9) is:

\begin{equation}
\Delta E\sim h,\quad\quad h\rightarrow 0.
\end{equation}

\noindent This means that
the asymptotic functional dependence of the resonance width is the
same as that of the SL width in frequency ranges beyond the
resonance. So, the functional dependence of the SL in the resonance
range is of the same type as beyond it, being $\propto h$. At the
same time, a ratio between the resonance width and $h$ may still be
a large number. Both these conclusions are in agreement with
computer simulations. Thus, for the archetypal example of the
Duffing oscillator subject to the dipole time-periodic perturbation
\cite{treshev,soskin2000}, the ratio $\Delta E/h$ approaches in the
limit $h\rightarrow 0$ the constant value approximately equal to 20:
see Fig. 3(b) in \cite{soskin2000}. For another archetypal example,
namely a pendulum in the wave of the same spatial period
\cite{Zaslavsky:1991,lichtenberg_lieberman,zaslavsky:1998,zaslavsky:2005},
our recent simulations for the parameters exploited in
\cite{Zaslavsky:1991,zaslavsky:1998,zaslavsky:2005} have shown that
$\Delta E/h\stackrel{h\rightarrow 0}{\longrightarrow}{\rm
const}\approx 50$.

\end{document}